\documentstyle[preprint,aps,epsfig]{revtex}

\def\be{\begin{equation}}                    
\def\de{\end{equation}}

\def\f{\frac}
\def\me{{\bf M}_d\cdot {\bf e}}

\begin{document}
\draft

\title{Nonlinear ac responses of  electro-magnetorheological fluids}
\author{J. P. Huang$^{1,2}$ and K. W. Yu$^2$}
\address{$^1$Department of Physics, The Chinese University of Hong Kong, Shatin, NT, Hong Kong \\
$^2$Max Planck Institute for Polymer Research, Ackermannweg 10, 55128, Mainz, Germany }

%\footnote{Correspondence address.}, and\\
%Department of Physics, The Chinese University of Hong Kong,
% Shatin, NT, Hong Kong}
\maketitle

\begin{abstract}

We  apply a Langevin model to investigate the nonlinear ac responses of electro-magnetorheological (ERMR) fluids under the application of two crossed dc magnetic ($z$ axis) and electric ($x$ axis) fields and a probing ac  sinusoidal magnetic field. We focus on the influence of the magnetic fields which can yield  nonlinear behaviors inside the system due to the particles with a permanent magnetic dipole moment. 
 Based on a perturbation approach, we extract the harmonics  of the magnetic field and  orientational magnetization  analytically. To this end, we find that the harmonics are sensitive to  the degree of anisotropy of the structure as well as the field frequency. 
Thus, it is possible to real-time monitor the structure transformation of ERMR fluids by detecting the nonlinear ac responses.

\end{abstract}
\newpage

\section{introduction}

Electro-magnetorheological (ERMR) fluids~\cite{Tao98,Sheng99} behave as both electrorheological (ER) fluids~\cite{Win49,Halsey,ud,GuJCP} and magnetorheological (MR) fluids~\cite{book,Bossis97,Martin03}, and are, in general, particle suspensions where the particles have large dielectric constants and permanent magnetic moments. In fact, there exist a variety of particles which can be polarized by both electric field and magnetic field~\cite{eg2}.  For instance, there is one candidate for this system, namely, a suspension of titanium-coated iron particles in a silicon oil~\cite{eg}.   As the external electric field or magnetic field exceeds a threshold, ER or MR fluids turn into a semi-solid, the ground state of which is a body-centered-tetragonal (bct) lattice. For an ERMR solid, one~\cite{Tao98}  proposed
that a structure transformation of the ERMR solid from the bct
lattice to some other lattices can appear when a magnetic field is
simultaneously applied perpendicular to the electric field. To one's interest,   such a structure transformation from 
bct  to  face-centered-cubic   lattice was experimentally observed
 as the ratio between the magnetic and the electric fields exceeded a minimum value~\cite{Sheng99}. Recently, one of the present authors showed that an
alternative structure transformation from the bct structure to the
fcc can appear under the application of electric fields
only~\cite{Lo01}.

Electric or magnetic fields of high strength applied to the ERMR solid produce a nonlinearity in the dependence of the polarization or magnetization on the field strength. Consequently, in the presence of an ac  electric  or magnetic field, the electrical or magnetic response will in general consist of ac fields at frequency of the higher-order  harmonics~\cite{C1,C2,C3,Hui04}. A convenient method of probing the nonlinear characteristics of the
composite is to measure the harmonics of the nonlinear
polarization (or magnetization) under the application of a sinusoidal ac
field~\cite{Kling98}. In this case,  the strength of the
nonlinear polarization or magnetization should be reflected in the magnitude of the
harmonics. For extracting such harmonics, the perturbation
approach~\cite{Gu00,PRE1} and self-consistent
method~\cite{PRE1,Wan01} can be used. To the best of our knowledge, no work has been done on the nonlinear ac responses of ERMR fluids.

%1-1

For the present system under consideration, the nonlinearity can be caused to appear by two effects, namely, normal saturation and anomalous saturation. In detail, the normal saturation arises from the higher terms of the Langevin function at large field intensities~\cite{Bot}. In contrast, the anomalous saturation results from  the equilibrium between entities with higher and lower dipole moments which is shifted under the influence of the field~\cite{Bot}. Our formalism will hold for the coupling between the normal saturation and anomalous saturation.

In the present paper, to investigate the structural effect  on
the nonlinear ac responses of ERMR solids, we shall apply a Langevin model to derive the orientational magnetization as well as the magnetic field inside the ERMR solid.
In this connection, the perturbation approach will be used to 
to extract the  harmonics of the magnetic field and  orientational magnetization.  

%To this end, it is shown that the fundamental and
%third-order harmonic ac responses are sensitive to the degree of
%anisotropy of the ER solid. Thus, by measuring the AC responses,
%it is possible to perform a real-time monitoring of the structure
%transformation of the ER solids.

This paper is organized as follows. In Sec.~II, we  apply the
Langevin model  to derive the orientational magnetization as well as the magnetic field inside the ERMR slid, and  calculate the effective permeability of the
ERMR solid by using a generalized  Clausius-Mossotti equation. Also, based on the perturbation approach, we  extract the harmonics of the magnetic field and orientational magnetization analytically.
In Sec.~III, we 
numerically investigate these harmonics as a function of the degree of anisotropy of the structure, as well as the frequency of the ac magnetic field. This is followed by a
discussion and conclusion in Sec.~IV.
\section{Formalism}

\subsection{Nonlinear characteristics}

Under the application of a strong magnetic field along $z$ axis, a nonlinear characteristic can appear in the ERMR fluid, due to the particles with a permanent magnetic dipole moment $p_0 .$ Accordingly, the dependence of the magnetic induction  ${\bf B}$ 
on the field ${\bf H}_0$ will be nonlinear~\cite{Bot} 
\be
{\bf B}=\mu_e {\bf H}_0+4 \pi \chi H_0^2{\bf H}_0,\label{DE2}
\de
where $\chi$ and  $\mu_e$ stand for the nonlinear susceptibility and effective permeability for the longitudinal field case, respectively.
In this case, the  effective permeability  $\mu_e$ is determined by the generalized  Clausius-Mossotti equation~\cite{Lo01}
\be
\frac{g_L(\mu_e-\mu_2)}{\mu_2+g_L(\mu_e-\mu_2)} = \frac{4\pi}{3}N_1\left (\alpha_1+
\frac{p_0^2}{3k_BT}\frac{1}{1+i 2\pi f\tau_1}\right ),\label{CME}
\de
where $\mu_2$ represents the permeability of the host fluid, $N_1$ the number density of the particles, $k_B$ the Boltzmann constant, $T$ the absolute temperature, $f$ the frequency of the applied magnetic field,  $\tau_1$ the relaxation time of the particles, and $\alpha_1$ the magnetizability of the particles. In Eq.~(\ref{CME}),  the longitudinal demagnetizing factor  $g_L$  deserves a thorough consideration. 
%In this case, the
%permanent  magnetic dipoles form a permanent magnet, and then
% there is a demagnetizing field,
%namely, the field inside the permanent magnet is opposite to the field
%outside, so that $\oint {\bf H}_0\cdot {\bf dl} = 0$ for any magnetic closed circuit. 
For an
isotropic array of magnetic dipoles, the demagnetizing factor will be
diagonal with the diagonal element $g_L=1/3 .$ However, in an anisotropic array like ERMR solids, the demagnetizing factor can still be diagonal, but it  deviates from $1/3$.  In fact, the degree of anisotropy of the system is just measured by how $g_L$ is deviated from $1/3 .$ It is worth noting that $g_L\le 1/3$ in the present longitudinal field case. Furthermore, there is a sum rule for
the factors, $g_L+2g_T =1$~\cite{PRE1,Land8}, where $g_T$ denotes the transverse demagnetizing factor. Such factors were measured by means of computer simulations~\cite{Martin98,Martin2}.  Thus, to investigate the anisotropic structural information of the array,  we have to modify
the Clausius-Mossotti equation accordingly by including the demagnetizing factor. The substitution of $g_L(=g_T)=1/3$ into Eq.~(\ref{CME}) yields the usual (isotropic) Clausius-Mossotti equation, which does not include the particle-particle interaction. In fact, when we studied the field-induced structure transformation in ER solids, we  developed the generalized Clausius-Mossotti    equation~\cite{Lo01} by introducing a local-field factor $\beta'$ which reflects the particle-particle interaction between the particles in a lattice. In detail, the generalized Clausius-Mossotti    approach  [Eq.~(\ref{CME})] is a self-consistent determination of the local  field due to a lattice of dipole moments. That is, Eq.~(\ref{CME})  should be expected to include the particle-particle interaction, and the degree of the particle-particle interaction depends on how much $g_L$ deviates from $1/3$ (note that  $g_L=\beta'/3$). Indeed our numerical results will also show that as $g_L$ deviates from $1/3$ more (namely, more and more particle chains are formed, and the particle-particle interaction becomes more and more strong, too), the obtained harmonics become more large accordingly, see Figs.~1-2. These harmonics just reflect the magnitude of the nonlinear susceptibility, as expected.

In Eq.~(\ref{CME}), the term $p_0^2/(3k_BT)$ results from the average contribution of the permanent magnetic dipole moment to the average value of the work required to bring a particle into the field $H_0$. More precisely, the mean value of the component of the permanent dipole moment in the direction of the field is given by
\begin{equation}
p_0 L(\gamma) = \frac{p_0^2}{3k_BT} H_0,
\end{equation}
with $\gamma = p_0 H_0/(k_BT).$ That is, we set the Langevin function
\begin{equation}
L(\gamma)=\gamma/3.\label{linear-L}
\end{equation}
 Regarding this linear relation in use, we should make some remarks. In the present work, we shall adopt the perturbation approach~\cite{Gu92}, which is suitable for weak nonlinearity. In this perturbation approach, it is well established that the effective third-order nonlinear susceptibility can be calculated from the linear field~\cite{Stroud88}, while the effective higher-order  nonlinearity must depend on
the nonlinear  field~\cite{Gu92}.  Alternatively, we could adopt the self-consistent method~\cite{PRE1}, but the perturbation approach appears to be more convenient for analytic expressions~\cite{PRE1}. Thus, for focusing on (weak) third-order nonlinearity, it suffices  to use the Clausius-Mossotti equation [Eq.~(\ref{CME})] by taking into account of  the linear relation [Eq.~(\ref{linear-L})] only. Due to the same reason, the contribution from the  nonlinear field will  be omitted throughout the paper.

%--

In view of Eq.~(\ref{DE2}), the desired 
field-dependent incremental permeability $\mu_H$ is given by~\cite{Bot}
\be
\mu_H=\f{\partial B}{\partial H}_0=\mu_e+12 \pi \chi H_0^2.
\de
Then, the nonlinear magnetic effect is characterized by $\triangle \mu/H_0^2$~\cite{Bot}
\be
\f{\triangle \mu}{H_0^2}=\f{\mu_H-\mu_e}{H_0^2}=12\pi \chi.\label{def}
\de

%\section{Formalism}

%Based a recent work [J. P. Huang, K. W. Yu, and M. Karttunen, Preprint: cond-mat/0403019],  the nonlinear dielectric increment is given by
%\begin{equation}
%\f{\triangle \mu}{H_0^2}
% = \f{18\pi}{5(kT)^3}\f{\mu_e^4}{(2\mu_e+\mu_{\infty})^2(2\mu_e^2+\mu_{\infty}^2)}\left(\f{\mu_{\infty}+2\mu_2}{3\mu_2}\right)^4\phi,\label{non}
%\end{equation}
%where
%$$
%\phi = p_0^4(-5VN_2^2+5N_2n_2-2N_2).
%$$
 
%In this equation, we neglect the 
%correlation between the orientations of the particles.

In the present work, the magnetization can be split up to two parts, i.e., the induced magnetization due to the magnetizability of the particles and the orientation magnetization $M_{or}$ due to the alignment of permanent magnetic dipole moments inside the particles. Let us now consider the orientation magnetization $M_{or}$ of a  
sphere of volume $V$, containing $n_1$ particles
 with a permanent magnetic dipole moment ${\bf p}_d$ [see Eq.~(\ref{Md2})] 
embedded in a continuum with permeability
 $\mu_{\infty}$ [see Eq.~(\ref{infty})]. 
  The sphere is surrounded by an infinite medium with the same 
macroscopic properties
  as the sphere. 

The average component in the direction of the field of the  magnetic dipole moment due the the dipoles  in the sphere 
$\langle {\bf M}_d\cdot {\bf e}\rangle =V M_{or}$ is given by:
\be
\langle {\bf M}_d\cdot {\bf e}\rangle =V M_{or}=\f{\int {\rm d}X^{n_1}{\bf 
M_d}\cdot {\bf e}e^{-u/k_BT}}{\int {\rm d}X^{n_1}e^{-u/k_BT}},\label{Md}
\de
where ${\bf e}$ denotes the unit vector in the direction of the external 
field, and $X$ stands for the set of position and orientation 
variables of all particles. Here $u$ is the 
energy related to the dipoles in the sphere, and it consists of three parts:
the energy of the dipoles in the external field $u_{de}$, the magnetostatic interaction energy of the dipoles $u_{mi}$, the non-magnetostatic interaction energy between the dipoles $u_{nmi}$ which is 
responsible for the short-range correlation between orientations and 
positions of the dipoles.    Regarding the particle-particle interaction energy, Eq.~(\ref{CME}) should be expected to contain both $u_{mi}$  and  $u_{nmi}$ (at least to some extent) as $g_L$ is not equal to $1/3$. Nevertheless, for the numerical calculations in the next section, Eqs.~(\ref{ab-1})~and~(\ref{ab-2})  will be used which  mean that  $u_{nmi}$ is omitted due to the predominant external magnetic field.      In Eq.~(\ref{Md}), ${\bf M}_d$ is given by 
\be
{\bf M}_d = 
\sum_{i=1}^{n_1}({\bf p}_d)_i,\label{Md2}
\de
with ${\bf p}_d = {\bf p}_0(\mu_{\infty}+2\mu_2)/3\mu_2$, 
where $\mu_{\infty}$  represents the permeability at frequencies 
at which the permanent dipoles  
cannot follow the changes of the field but where the atomic and the 
electronic magnetization are still the same as in the static field.
Therefore, $\mu_{\infty}$ is the permeability characteristic for 
the induced magnetization. In practice,  $\mu_{\infty}$ can be expressed in the expression containing an intrinsic dispersion,
\be
\mu_{\infty} = \mu_{\infty}(0)+\frac{\Delta\mu}{1+i f/f_c},\label{infty}
\de
where $\mu_{\infty}(0)$ is the high-frequency limit permeability, and $\Delta\mu$ stands for the magnetic dispersion strength with a characteristic frequency $f_c .$
%In practice, the atomic polarization can often be neglected, and hence 
%one may set $\mu_{\infty}$ to be the 
%square of the refractive index, which is readily obtained experimentally. 

The external field in this model is equal 
to the field within
 a spherical cavity filled with a continuum of permeability
 $\mu_{\infty}$, while the cavity is situated in a medium with 
permeability $\mu_e$. This field is called Fr\"{o}hlich field $H_F$, as given by the Fr\"ohlich model~\cite{Bot,Fro}
\be
H_F=\frac{3\mu_e}{2\mu_e+\mu_{\infty}}H_0+\frac{12\pi\chi 
\mu_{\infty}}{(2\mu_e+\mu_{\infty})^2}H_0^3.\label{Field}
\de
In this equation, higher-order terms have been omitted. Regarding the derivation of the nonlinear Fr\"{o}hlich field [Eq.~(\ref{Field})], here we would like to add some comments. For the spherical cavity under consideration, the linear cavity field (i.e., linear Fr\"{o}hlich field) $H_{F}^{{\rm lin}}$ can be easily obtained by solving the usual electrostatic equation. That is, it should be the first term of the right-hand side of Eq.~(\ref{Field}). Next, to obtain the present nonlinear Fr\"olich field $H_F$, following Ref.~\cite{Bot} we took one step forward to develop  $H_{F}^{{\rm lin}}$   in a Taylor series around $\mu_e$ and neglecting terms in $H_0^4$ and higher powers of $H_0$.  It is worth noting that the higher powers of $H_0$ can also contribute to the resulting harmonics. However, this contribution is small enough to be neglected.

Now we use
\be
\frac{\partial u}{\partial H_F}=-{\bf M}_d\cdot {\bf e}.
\de
Then, we obtain
\begin{eqnarray}
\langle {\bf M}_d\cdot {\bf e}\rangle |_{H_F=0}&=& 0,\\
\frac{\partial}{\partial_{H_F}}\langle {\bf M}_d\cdot {\bf 
e}\rangle &=&\frac{1}{k_BT}[\langle ({\bf M}_d\cdot {\bf 
e})^2\rangle -\langle \me\rangle^2 ],\\
\frac{\partial}{\partial_{H_F}}\langle {\bf M}_d\cdot {\bf 
e}\rangle |_{H_F=0}&=&\f{1}{k_BT}\langle M_d^2\rangle _0,\\
\f{\partial^3}{\partial H_F^3}\langle \me\rangle &=&\f{1}{(k_BT)^3}[\langle 
(\me)^4\rangle -3\langle (\me)^3\rangle \nonumber\\
        & &\langle \me\rangle +6\langle (\me)^2\rangle \langle \me\rangle ^2-3\langle (\me)^2\rangle^2 - \nonumber\\
       & & 6\langle \me\rangle ^4+6\langle \me\rangle^2 \langle (\me)^2\rangle-\langle \me\rangle \langle (\me)^3\rangle ],\\
\f{\partial^3}{\partial 
H_F^3}\langle \me\rangle |_{H_F=0}&=&\f{1}{15(k_BT)^3}[3\langle M_d^4\rangle _0-5\langle M_d^2\rangle ^2_0].
\end{eqnarray}
Note the subscript $0$ indicates the absence of the field. 

To express the macroscopic saturation behavior in terms of microscopic quantities, one should take into account the
higher derivatives of the average 
moment such that
\begin{eqnarray}
\langle \me\rangle &=&\f{\partial\langle \me\rangle }{\partial 
H_F}|_{H_F=0}H_F+\f{1}{6}\f{\partial^3\langle \me\rangle }{\partial 
H_F^3}|_{H_F=0}H_F^3\nonumber\\
&=&\frac{\mu_e}{2\mu_e+\mu_{\infty}}\f{\langle M_d^2\rangle_0}{k_BT}H_0+\f{4\pi \chi 
\mu_{\infty}}{(2\mu_e+\mu_{\infty})^2}\f{\langle M_d^2\rangle _0}{k_BT}H_0^3\nonumber\\
& 
&+\f{27\mu_e^3}{(2\mu_e+\mu_{\infty})^3}\f{3\langle M_d^4\rangle _0-5\langle M_d^2\rangle _0^2}{90(k_BT)^3}H_0^3,
\end{eqnarray}
where higher-order terms than 3rd have been neglected.

In addition, we have a general relation
\be
\langle \me\rangle =VM_{or}=\f{\mu_e-\mu_{\infty}}{4\pi}VH_0+\chi VH_0^3.
\de
In view of Eq.~(\ref{def}) and the terms in $H_0^3$ and $H_0$, we obtain
\be
\f{\triangle 
\mu}{H_0^2}=\frac{18\pi}{5(k_BT)^3}\f{\mu_e^4}{(2\mu_e+\mu_{\infty})^2(2\mu_e^2+\mu_{\infty}^2)}[3(\langle M_d^4\rangle _0/V)-5(\langle M_d^2\rangle _0^2/V)].\label{DelE}
\de
On the other hand, we may write
\begin{eqnarray}
\f{\langle M_d^2\rangle _0}{V}&=&\left(\f{\mu_{\infty}+2\mu_2}{3\mu_2}\right)^2[\frac{n_1}{V} 
p_0^2\sum_{j=1}^{n_1}\langle \cos \theta_{ij}\rangle ],\label{def-1} \\
\f{\langle M_d^4\rangle _0}{V}&=&\left(\f{\mu_{\infty}+2\mu_2}{3\mu_2}\right)^4[\frac{n_1}{V} 
p_0^4\sum_{j=1}^{n_1}\langle \cos 
\theta_{ij}\sum_{r=1}^{n_1}\sum_{s=1}^{n_1}\cos \theta_{rs}\rangle ].\label{def-2}
 \end{eqnarray}
In the right-hand sides of both Eqs.~(\ref{def-1})~and~(\ref{def-2}), the $\langle\cdots \rangle$'s denote
\begin{equation}
\langle\cdots \rangle = \int \frac{\int {\rm d}X^{n_1-i} \cdots e^{-u/k_BT}}{\int {\rm d}X^{n_1}  e^{-u/k_BT}} {\rm d}X^i.
\end{equation}
 Hence the nonlinear magnetic increment $\triangle \mu/H_0^2$ [Eq.~(\ref{DelE})]  can be 
explicitly expressed.

\subsection{Nonlinear magnetization and high-order harmonics}

\subsubsection{Longitudinal field}

For the longitudinal field case, there is $H_0(t)=H_{{\rm dc}}+H_{{\rm ac}}(t)=H_{{\rm dc}}+H_{{\rm ac}}\sin \omega t$ along $z$ axis, with $\omega=2\pi f .$ Here $H_{{\rm dc}}$ denotes the dc field which induces  the anisotropic structure in the ERMR solid, and $H_{{\rm ac}}(t)$ stands for a sinusoidal ac field. Then, the
Fr\"{o}hlich field is a superposition of odd- and even- order harmonics such that
\be
 H_F = H_F^{{\rm (dc)}}+H_{\omega}\sin\omega t+ H_{2\omega}\cos 2\omega t + H_{3\omega}\sin 3\omega t+\cdots.\label{Lo-E}
\de
Accordingly, the orientational magnetization contains harmonics as
\be
M_{{\rm or}}=M_{{\rm or}}^{{\rm (dc)}}+M_{\omega}\sin\omega t+M_{2\omega}\cos 2\omega t  +M_{3\omega}\sin 3\omega t+\cdots.
\de
where the harmonics   ($H_{\omega},$ $H_{2\omega},$ $H_{3\omega},$  $M_{\omega},$ $M_{2\omega},$ and $M_{3\omega}$) and the dc components  ($H_F^{{\rm (dc)}}$ and  $M_{{\rm or}}^{{\rm (dc)}}$)  can be expressed as 
\begin{eqnarray}
H_F^{{\rm (dc)}} &=& H_{{\rm dc}}I_1+\frac{3}{2}H_{{\rm ac}}^2H_{{\rm dc}}I_3+H_{{\rm dc}}^3I_3,\\
H_{\omega} &=& H_{{\rm ac}}I_1+\frac{3}{4}H_{{\rm ac}}^3I_3+3H_{{\rm ac}}H_{{\rm dc}}^2I_3,\\
H_{2\omega} &=& -\frac{3}{2}H_{{\rm ac}}^2H_{{\rm dc}}I_3,\\
H_{3\omega} &=& -\frac{1}{4}H_{{\rm ac}}^3I_3,\\
M_{{\rm or}}^{{\rm (dc)}} &=&  H_{{\rm dc}}J_1+\frac{3}{2}H_{{\rm ac}}^2H_{{\rm dc}}\chi+H_{{\rm dc}}^3\chi,\\
M_{\omega} &=& H_{{\rm ac}}J_1+\frac{3}{4}H_{{\rm ac}}^3\chi+3H_{{\rm ac}}H_{{\rm dc}}^2\chi,\\
M_{2\omega} &=& -\frac{3}{2}H_{{\rm ac}}^2H_{{\rm dc}}\chi,\\
M_{3\omega} &=& -\frac{1}{4}H_{{\rm ac}}^3\chi,\label{Lo-M3}
\end{eqnarray}
with $I_1=3\mu_e/(2\mu_e+\mu_{\infty}),$ $I_3=12\pi \chi \mu_{\infty}/(2\mu_e+\mu_{\infty})^2,$ and $J_1=(1/4\pi)(\mu_e-\mu_{\infty}).$  In the above derivation, we  used two identities, i.e. $\sin^2 \omega t=(1-\cos 2\omega t)/2$ and $\sin^3 \omega t = (3/4)\sin\omega t - (1/4) \sin 3\omega t$.

\subsubsection{Transverse field}

For the transverse field case, only a sinusoidal electric field
$
H_0(t)=H_{{\rm ac}}\sin\omega t\label{E0t}
$
 is applied along $x$ axis. In this connection, the Fr\"{o}hlich   field is a superposition of odd-order harmonics 
\be
H_F =H_{\omega}\sin\omega t+H_{3\omega}\sin 3\omega t+\cdots.\label{Tr-E}
\de
Accordingly, the  orientational magnetization  contains harmonics as
\be
M_{{\rm or}}=M_{\omega}\sin\omega t+M_{3\omega}\sin 3\omega t+\cdots.\label{Tr-M}
\de
In Eqs.~(\ref{Tr-E})~and~(\ref{Tr-M}),  the harmonics  $H_{\omega},$ $H_{3\omega},$  $M_{\omega},$ $M_{3\omega}$ can be expressed as

\begin{eqnarray}
H_{\omega} &=& H_{{\rm ac}}I_1'+\frac{3}{4}H_{{\rm ac}}^3I_3',\\
H_{3\omega} &=& -\frac{1}{4}H_{{\rm ac}}^3I_3',\\
M_{\omega} &=& H_{{\rm ac}}J_1'+\frac{3}{4}H_{{\rm ac}}^3\chi',\\
M_{3\omega} &=& -\frac{1}{4}H_{{\rm ac}}^3\chi'.\label{Tr-M3}
\end{eqnarray}
with $I_1'=3\mu_e'/(2\mu_e'+\mu_{\infty}),$ $I_3'=12\pi \chi' \mu_{\infty}/(2\mu_e'+\mu_{\infty})^2,$ and $J_1'=(1/4\pi)(\mu_e'-\mu_{\infty}),$ where $\mu_e'$ and $\chi'$ denote the effective permeability and nonlinear susceptibility in the transverse field case.
  For the transverse field case, $\chi'$ is induced to appear by the moderate ac magnetic field only. Thus, in view of Eqs.~(\ref{Md})~and~(\ref{Md2}), we find  $\chi'\ll \chi$ ($\chi$ is caused to appear by both the moderate ac and high dc magnetic fields in the longitudinal field case, as mentioned above). As a result, the third-order harmonics of the Fr\"{o}hlich  field and  orientational magnetization are much smaller in the transverse field case than in the longitudinal field case.  In this regard, for the following numerical calculations, we shall focus on the longitudinal field case only.

\section{numerical results}

%1-3

%In practice, the increasing magnetic
%field can be predominant in overcoming the dipolar interactions between
%the magnetic particles, leading to the nonlinear effect. 
%Thus, in  the numerical calculations we shall neglect the 
%correlation between the orientations of the particles.  In this connection, we have

Now we are in a position to do some numerical calculations. For the numerical calculation, we use
\begin{eqnarray}
\sum_{j=1}^{n_2}\langle \cos \theta_{ij}\rangle &=&1 ,\label{ab-1} \\
\sum_{j=1}^{n_2}\langle \cos \theta_{ij}\sum_{r=1}^{n_2}\sum_{s=1}^{n_2}\cos 
\theta_{rs}\rangle &=&\f{1}{3}(5 n_1-2).\label{ab-2}
\end{eqnarray}
Eqs.~(\ref{ab-1})~and~(\ref{ab-2}) imply that the correlation between the dipole moments of the particles and in turn the anomalous saturation are neglected in the sense that the current anomalous saturation is much more weak. This is reasonable because, in the presence of a magnetic field, the permanent magnetic moments of the particles are easily directed along the field. Therefore, the equilibrium between the higher dipole moment (of particle chains) and the lower dipole moment (of particle chains) might not be able to predict significant nonlinearity, when compared to the normal saturation. 

Next, we  take one step forward to obtain the nonlinear magnetic increment
\begin{equation}
\f{\triangle \mu}{H_0^2}
 = \frac{-36\pi N_1p_0^4}{5(k_BT)^3}\f{\mu_e^4}{(2\mu_e+\mu_{\infty})^2(2\mu_e^2+\mu_{\infty}^2)}
\left(\f{\mu_{\infty}+2\mu_2}{3\mu_2}\right)^4.\label{non}
\end{equation}

 For numerical calculations, take the following parameters:  $H_{{\rm dc}}=100\,$Oe, $H_{{\rm ac}}=1\,$Oe,   $\mu_{\infty}(0)=1.1 ,$ $\Delta\mu=8 ,$ $f_c =4.8\times 10^3\,$Hz, $\mu_2 = 1 ,$  $p_0=10^{-13}\,$emu, $\tau_1=4.8\times 10^{-7}\,$s, and $\alpha_1=10^{-10}\,$cm$^{-3} .$ In addition, the radius of particle is taken to be $5\,$$\mu$m, and the volume fraction of the particles is $0.2 .$

Fig.~1 shows the fundamental, second- and third- order harmonics of the Fr\"{o}hlich field as a function of the field frequency, for various $g_L$ in the longitudinal field case. In this figure, a peak is observed always due to the existence of an intrinsic dispersion [Eq.~(\ref{infty})]. In particular, as the system changes from isotropic case ($g_L=1/3$) to anisotropic ($g_L\ne 1/3$) because of the appearence of the particle chains, the harmonics of the field can be changed accordingly. In detail,  stronger anisotropy  (namely, decreasing the longitudinal demagnizing factor $g_L$) leads to larger harmonics, especially in the low-frequency region.   Similar effect can be shown in Fig.~2 where  the harmonics of the orientational magnetization in the longitudinal field case are investigated for various $g_L$ as well. However, the fundamental harmonic of the orientation magnetization [see Fig.~2(a)] behaves in a different way from that of the Fr\"{o}hlich field  [see Fig.~1(a)]. In detail, as the frequency increases, this fundamental harmonic  decreases first, then increases, and after reaching a peak they decreases again. 

%Similarly, in contrast to the isotropic system, the harmonics of orientational magnetization can also be changed  by the  anisotropy of the structure.  Also, stronger anisotropy leads to larger harmonics. 

For the transverse field case, it could be concluded that as the demagtizing factor $g_L$ decreases, both the fundamental and  third-order harmonics of the  Fr\"{o}hlich field and  orientation magnetization are caused to decrease accordingly, which is just opposite with those obtained from the longitudinal field case (no figures shown here). The reason is that there is a sum rule between $g_L$ and $g_T$,  $g_L+2g_T=1 .$

In a word, for the longitudinal field case, besides the odd-order harmonics, the even-order harmonics are also induced to appear due to the coupling between the applied dc and ac magnetic fields along $z$ axis [see Figs.~1~and~2, or Eqs.~(\ref{Lo-E})$\sim$(\ref{Lo-M3})] even though only the cubic nonlinearity [Eq.~(\ref{DE2})] is considered  due to the virtue of symmetry of the system. On the other hand, for the transverse field case (i.e., there is only a single ac magnetic field applied along $x$ axis), in view of the cubic nonlinearity  [Eq.~(\ref{DE2})]  of interest, only the odd-order harmonics are induced, as already  predicted by Eqs.~(\ref{Tr-E})$\sim$(\ref{Tr-M3}).  
%In a word, for the transverse field case, only the odd-order harmonics  appear [see Eqs.~(\ref{Tr-E})$\sim$(\ref{Tr-M3})]. In contrast, for the longitudinal field case, both the  odd- and even- order harmonics can be induced to appear [see Eqs.~(\ref{Lo-E})$\sim$(\ref{Lo-M3})].

In addition, even though there is no particle-particle interaction (i.e., $u_{mi}=u_{nmi}=0$)  as $g_L=1/3$, the nonlinear behavior due to the normal saturation could still be induced to occur because of the presence of external fields, i.e.,  $u_{de}\ne 0$. This is the reason why the harmonics shown in Figs.~1-2 are nonzero at $g_L=1/3$.

From Figs.~1~and~2, we find that the second-order harmonics  of the Fr\"{o}hlich field and  orientational magnetization  are of three orders of magnitude larger than the corresponding third-order harmonics. Thus, to monitor the structure transformation of ERMR solids,  it is more attractive to detect the second-order harmonics than the third-order.

Finally, we display the temperature effect on the harmonics of the Fr\"{o}hlich field and  orientational magnetization in Fig.~3 and Fig.~4, respectively. It is shown that decreasing the temperature $T$ causes all the harmonics to increase because of the change in the Fr\"{o}hlich field.

\section{Discussion and conclusion}

Here some comments are in order. Based on a Langevin model, we have investigated the nonlinear ac responses of ERMR solids which are subjected to a structural transformation.  In the present work, we have focused on the effect of the magnetic fields on the nonlinear ac responses. In fact, for the ERMR solid, the influence of electric fields can be investigated as well. However, the present Langevin model is invalid for this purpose because there is usually no permanent electric dipole moment inside the suspended particles, except for ferroelectric particles. To one's interest, when the particles own a nonlinear characteristic inside them, the electric-field effect on the nonlinear ac responses can still be discussed, based on the perturbation approach  and self-consistent method~\cite{PRE1} accompanying with an effective medium theory like the Maxwell-Garnett approximation.    

In this paper, the Fr\"ohlich field is actually an effective field, which is similar to the Lorentz local
field. The latter is only defined for induced dipole moments, while the
former is introduced to deal with permanent dipole moments.

%Throughout the paper, the nonlinear susceptibility $\chi$ is due to the saturation effects implied in the
%Langevin function. As the individual dipole moment experiences the effective
%field, the nonlinearity is just the product of the nonlinearity susceptibility
%and the square of the effective field.

The present consideration can be extended to ferrofluids, which are a suspension~\cite{ferro} containing ferromagnetic particles embedded in a carrier liquid. However, the particles in a ferrofluid possess a much smaller size than  in the ERMR solid. Also, the particles can form chains inside ferrofluids under the presence of a moderate magnetic field, without the need of a strong magnetic field. For ERMR solids, a strong magnetic field has to be used to induce the formation of particle chains.

Finally, since the particles in ERMR solids are located  very close, it is instructive to take into account the effect of multipolar interactions between the particles~\cite{Yu00,PRE2,Sun03} on the nonlinear ac responses.

To sum up,  we have applied  a Langevin model to investigate the nonlinear ac responses of ERMR solids. For the longitudinal field case, it has been shown that both even- and odd- order harmonics are induced to occur. In contrast, only the odd-order harmonics appear for the transverse field case. Moreover, these harmonics can be affected by the degree of anisotropy of the ERMR solid, as well as the field frequency. In particular, the second-order harmonics are of several orders of magnitude larger than the corresponding third-order. Thus, it is possible to real-time monitor the structural transformation of ERMR solids by detecting the nonlinear ac reponses.

\section*{Acknowledgments}

This work has been supported  by the Research Grants Council of
the Hong Kong SAR Government under Project No.~CUHK 403303, and in part by
the DFG under Grant No. HO 1108/8-3 (J.P.H.). We would like to thank Professor G. Q. Gu for useful discussions.

 \newpage

\newpage
\begin{figure}[h]
\caption{ Fundamental, second- and third- order harmonics of Fr\"{o}hlich field as a function of the field frequency, for various $g_L .$ Parameter: $T=298\,$K.  }
\end{figure}

%\newpage
\begin{figure}[h]
\caption{Fundamental, second- and third- order   harmonics of the orientational magnetization as a function of the field frequency, for various $g_L .$  Parameter: $T=298\,$K.  }
\end{figure}

\begin{figure}[h]
\caption{Same as Fig.~1, but for different temperatures. Parameter: $g_L=1/5 .$  }
\end{figure}

\begin{figure}[h]
\caption{Same as Fig.~2, but for different temperatures. Parameter: $g_L=1/5 .$   }
\end{figure}

%\newpage
%\begin{figure}[h]
%\caption{{\it Transverse field:} Fundamental and third-order harmonics of Fr\"{o}hlich field as a function of the field frequency, for various $g_L .$   }
%\end{figure}

%\newpage
%\begin{figure}[h]
%\caption{Same as Fig.~3, but for fundamental and third-order  harmonics of the orientational magnetization.  }
%\end{figure}

\newpage
\centerline{\epsfig{file=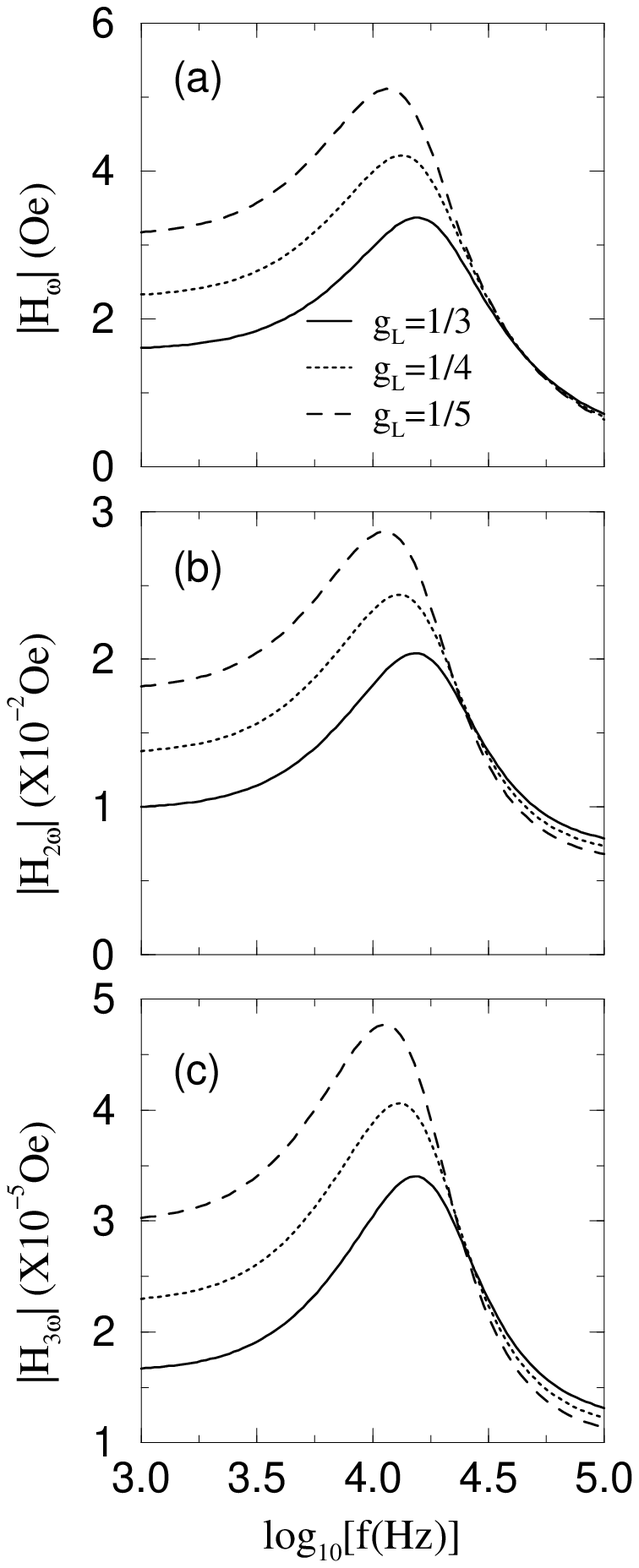,width=86mm}}
\centerline{Fig.~1/Huang and Yu}

\newpage
\centerline{\epsfig{file=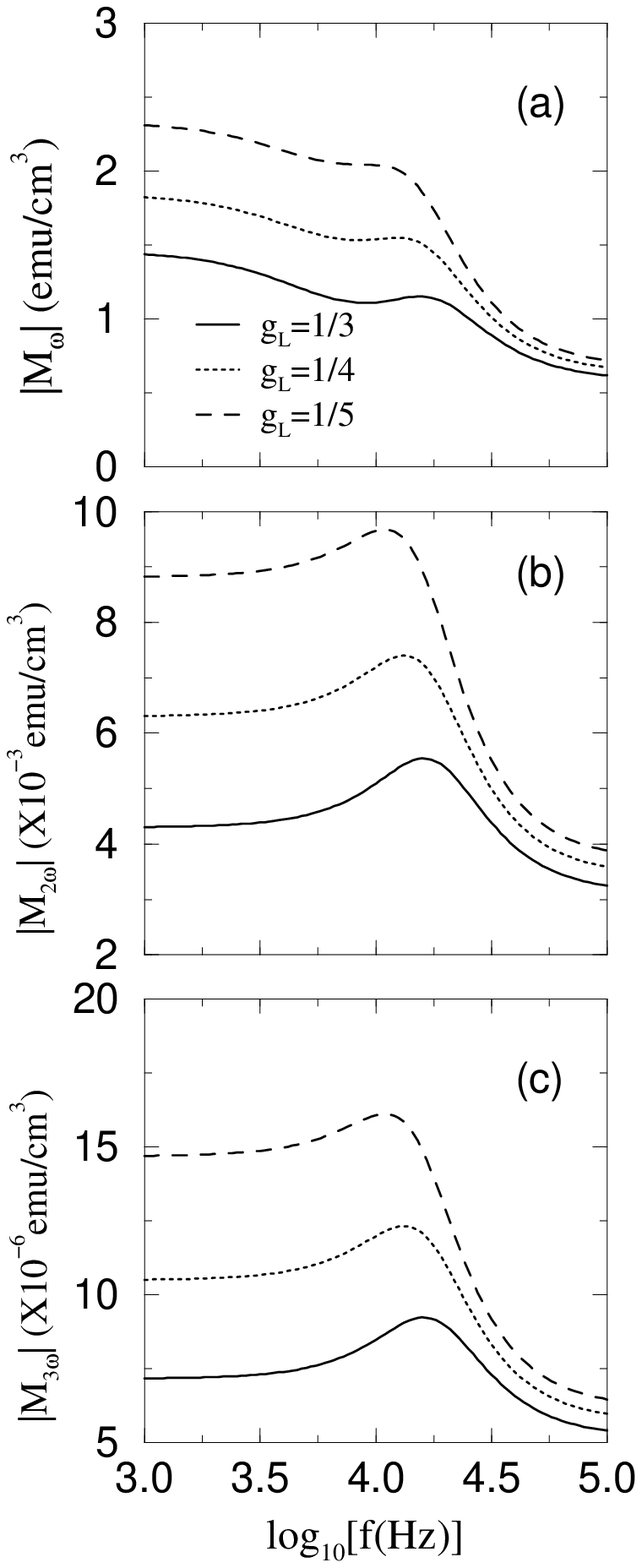,width=86mm}}
\centerline{Fig.~2/Huang and Yu}

\newpage
\centerline{\epsfig{file=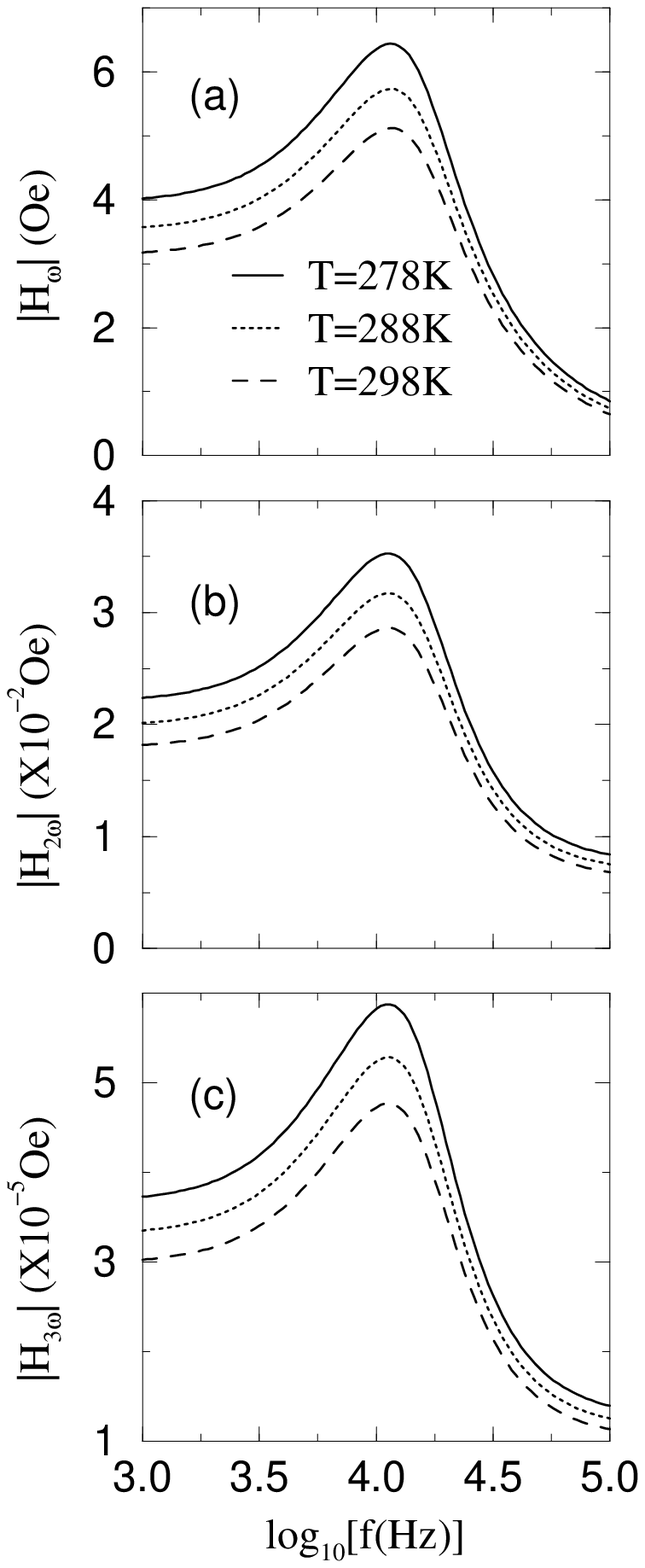,width=86mm}}
\centerline{Fig.~3/Huang and Yu}

\newpage
\centerline{\epsfig{file=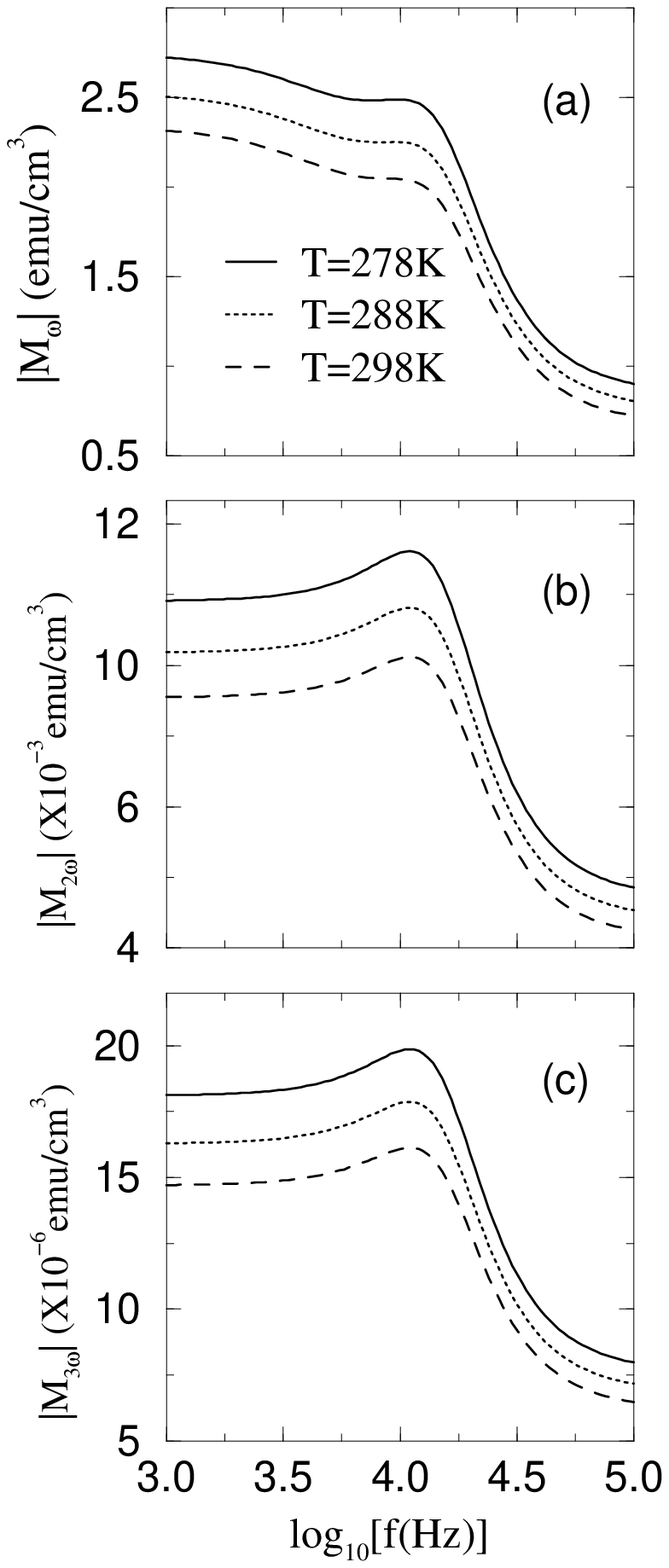,width=86mm}}
\centerline{Fig.~4/Huang and Yu}

%\newpage
%\centerline{\epsfig{file=fig3.eps,width=250pt}}
%\centerline{Fig.~3/Huang and Yu}
%\newpage
%\centerline{\epsfig{file=fig4.eps,width=250pt}}
%\centerline{Fig.~4/Huang and Yu}


\newpage

\begin{references}



\bibitem{Tao98} R. Tao and Q. Jiang, Phys. Rev. E {\bf 57}, 5761 (1998). 

%\bibitem{Sheng00}  P. Sheng {\it et al.}, Physica B {\bf 279} (2000) 168.

\bibitem{Sheng99} W. J. Wen, N. Wang, H. R. Ma, Z. F. Lin, W. Y. Tam, C. T. Chan, and P. Sheng, Phys. Rev. Lett. {\bf 82}, 4248 (1999).

%\bibitem{Sheng00} P. Sheng, W. Wen, N. Wang, H. Ma, Z. Lin, W. Y. Tam, and C. T. Chan, Physica B {\bf 279} (2000) 168.

\bibitem{Win49} W. M. Winslow, J. Appl. Phys. {\bf 20}, 1137 (1949).


\bibitem{Halsey} For a review, see T. C. Halsey, Science {\bf 258}, 761 (1992).

\bibitem{ud} U. Dassanayake, S. Fraden, and A. V. Blaaderen, J. Chem. Phys. {\bf 112}, 3851 (2000).

\bibitem{GuJCP} G. Q. Gu, K. W. Yu, and P. M. Hui, J. Chem. Phys. {\bf 116}, 10989 (2002).

\bibitem{book} V. I. Kordonsky and Z. P. Shulman, in {\it Electrorheological Fluids}, edited by J. D. Carlson, A. F. Sprecher, and H. Conrad (Technomic Publishing, Lancaster, Basel, 1991), pp. 437-444.

\bibitem{Bossis97} S. Cutillas and G. Bossis, Europhys. Lett. {\bf 40}, 465 (1997).

\bibitem{Martin03} S. Melle and J. E. Martin, J. Chem. Phys. {\bf 118}, 9875 (2003).




\bibitem{eg2} For example, see {\it Electrorheological Fluids}, edited by R. Tao (World Scientific, Singapore, 1992).

\bibitem{eg} K. Koyama, in {\it Electro-Rheological Fluids, Magneto-Rheological Suspensions and Associated Technology}, edited by W. A. Bullough (Word Scientific, Singapore, 1996), pp. 245-250.

\bibitem{Lo01}  C. K. Lo and K. W. Yu, Phys. Rev. E {\bf 64}, 031501 (2001).

\bibitem{C1} O. Levy, D. J. Bergman, and D. Stroud, Phys. Rev. E {\bf 52}, 3184 (1995).

\bibitem{C2} P. M. Hui and D. Stroud, J. Appl. Phys. {\bf 82}, 4740 (1997). 

\bibitem{C3} P. M. Hui, P. C. Cheung, and D. Stroud, J. Appl. Phys. {\bf 84}, 3451 (1998).

\bibitem{Hui04} P. M. Hui, C. Xu, and D. Stroud, Phys. Rev. B {\bf 69}, 014203 (2004).

\bibitem{Kling98} D. J. Klingenberg, MRS Bull. {\bf 23}, 30 (1998).

\bibitem{Gu00} G. Q. Gu, P. M. Hui, and K. W. Yu, Physica B {\bf
279}, 62 (2000).

 \bibitem{PRE1} J. P. Huang, J. T. K. Wan, C. K. Lo, and K. W. Yu,
Phys. Rev. E {\bf 64}, 061505(R) (2001).

% \bibitem{JAP1} J. P. Huang, L. Gao, and K. W. Yu, J. Appl. Phys. {\bf 93}, 2871 (2003).

%\bibitem{Phule98} P. P. Phul\'{e}, J. M. Ginder, MRS Bull. {\bf 23} (1998) 19.


\bibitem{Wan01} J. T. K. Wan, G. Q. Gu, and K. W. Yu, Phys. Rev. E
{\bf 63}, 052501 (2001).



\bibitem{Bot} 
C.\,J.\,F. B\"{o}ttcher, {\em Theory of electric polarization}, Vol.\,1, second edition 
(Elsevier, Amsterdam, 1993).

\bibitem{Land8}  L. D. Landau, E. M. Lifshitz, and L. P. Pitaevskii, {\it Electrodynamics of Continuous Media}, 2nd ed. (Pergamon press, New York, 1984), Chap.~II.

\bibitem{Martin98} J. E. Martin, R. A. Anderson, and C. P. Tigges, J. Chem. Phys. {\bf 108}, 3765 (1998).

\bibitem{Martin2} J. E. Martin, R. A. Anderson, and C. P. Tigges, J. Chem. Phys. {\bf 108}, 7887 (1998).

\bibitem{Gu92}  G. Q. Gu and K. W. Yu, Phys. Rev. B {\bf 46}, 4502 (1992).

\bibitem{Stroud88} D. Stroud and P. M. Hui, Phys. Rev. B {\bf 37}, 8719 (1988).

\bibitem{Fro} 
H. Fr\"{o}hlich, {\em Theory of dielectrics}, Oxford University 
Press, London 1958.

\bibitem{ferro} R. E. Rosensweig, {\it Ferrohydrodynamics} (Cambridge Univ. Press, Cambridge, 1985).

\bibitem{Yu00} K. W. Yu and J. T. K. Wan, Comput. Phys. Commun. {\bf 129}, 177 (2000).



\bibitem{PRE2} J. P. Huang, K. W. Yu, and G. Q. Gu, Phys. Rev. E {\bf 65}, 021401 (2002).

\bibitem{Sun03} H. Sun and K. W. Yu, Phys. Rev. E {\bf 67}, 011506 (2003).

\end{references}
\end{document}